\documentclass[preprint]{aastex}

\begin{document}

\title{Probing the Evolution of IR Properties of $z\sim6$ Quasars: 
{\it Spitzer} Observations}
\author{Linhua Jiang\altaffilmark{1}, Xiaohui Fan\altaffilmark{1},
  Dean C. Hines\altaffilmark{2}, Yong Shi\altaffilmark{1},
  Marianne Vestergaard\altaffilmark{1},
  Frank Bertoldi\altaffilmark{3}, W. N. Brandt\altaffilmark{4},
  Chris L. Carilli\altaffilmark{5}, Pierre Cox\altaffilmark{6},
  Emeric Le Floc'h\altaffilmark{1}, Laura Pentericci\altaffilmark{7},
  Gordon T. Richards\altaffilmark{8,9},
  George H. Rieke\altaffilmark{1}, Donald P. Schneider\altaffilmark{4},
  Michael A. Strauss\altaffilmark{8}, Fabian Walter\altaffilmark{10},
  and J. Brinkmann\altaffilmark{11}}
\altaffiltext{1}{Steward Observatory, University of Arizona,
  933 North Cherry Avenue, Tucson, AZ 85721}
\altaffiltext{2}{Space Science Institute, 4750 Walnut Street, Suite 205, 
  Boulder, CO 80301}
\altaffiltext{3}{Argelander Institute for Astronomy, University of Bonn,
  Auf dem H\"{u}gel 71, 53121 Bonn, Germany}
\altaffiltext{4}{Department of Astronomy and Astrophysics, Pennsylvania
  State University, 525 Davey Laboratory, University Park, PA 16802}
\altaffiltext{5}{National Radio Astronomy Observatory, P.O. Box 0, 
  Socorro, NM 87801}
\altaffiltext{6}{Institut de Radioastronomie Millimetrique, St. Martin
  d'Heres, F-38406, France}
\altaffiltext{7}{Dipartimento di Fisica, Universit\`{a} degli Studi Roma 3,
  Via della Vasca Navale 84, Roma, Italy}
\altaffiltext{8}{Department of Astrophysical Sciences, Peyton Hall, Princeton,
  NJ 08544}
\altaffiltext{9}{Department of Physics and Astronomy, The Johns Hopkins
  University, 3400 North Charles Street, Baltimore, MD 21218}
\altaffiltext{10}{Max-Planck-Institute for Astronomy, K\"{o}nigstuhl 17, 
  D-69117 Heidelberg, Germany}
\altaffiltext{11}{Apache Point Observatory, P.O. Box 59, Sunspot, NM 88349}

\begin{abstract}

We present $Spitzer$ observations of thirteen $z\sim6$ quasars using the 
Infrared Array Camera (IRAC) and Multiband Imaging Photometer for $Spitzer$ 
(MIPS). All the quasars except SDSS J000552.34$-$000655.8 (SDSS J0005$-$0006) 
were detected with high S/N in the four IRAC channels and the MIPS 
24$\mu$m band, while SDSS J0005$-$0006 was marginally detected in the IRAC 
8.0$\mu$m band, and not detected in the MIPS 24$\mu$m band. We find that 
most of these quasars have prominent emission from hot dust as evidenced by 
the observed 24$\mu$m fluxes. Their spectral energy distributions (SEDs) are 
similar to those of low-redshift quasars at rest-frame 0.15$-$3.5 $\mu$m, 
suggesting that accretion disks and hot-dust structures for these sources
already have reached maturity. However, SDSS J0005$-$0006 has an unusual SED 
that lies significantly below low-redshift SED templates at rest-frame 1 and 
3.5 $\mu$m, and thus shows a strong near-IR (NIR) deficit and no hot-dust 
emission. Type I quasars with extremely small NIR-to-optical flux ratios like 
SDSS J0005$-$0006 are not found in low-redshift quasar samples, indicating 
that SDSS J0005$-$0006 has different dust properties at high redshift.
We combine the $Spitzer$ observations with X-ray, UV/optical, 
mm/submm and radio observations to determine bolometric luminosities for 
all the quasars. We find that the four quasars with central black-hole mass
measurements have Eddington ratios of order unity.

\end{abstract}


\section{INTRODUCTION}

High-redshift quasars provide direct probes of the distant early universe
where the first generation of galaxies and quasars formed. In the last few
years, more than 20 luminous quasars at $z>5.7$ have been discovered by the
Sloan Digital Sky Survey \citep[SDSS;][]{yor00}; the most distant one is at 
$z=6.42$ \citep[e.g.][]{fan00,fan01,fan03,fan04,fan06}. The discovery of 
these luminous objects at $z\sim6$ reveals the existence of supermassive 
black holes (BHs) with masses higher than $10^9$M$_\sun$ 
\citep[e.g.][]{bar03,ves04} in the first Gyr of cosmic history. These 
quasars provide a unique high-redshift sample to answer a series of 
challenging questions: How did the first billion-solar-mass BHs appear less 
than 1 Gyr after the Big Bang? How did they grow with time? Did the physical
structure of quasars/AGNs evolve with time? Is the emission mechanism in 
quasars/AGNs the same at $z\sim6$ as at $z\sim0$? How were quasars
and starburst activity related in the earliest galaxies? 
What is the role the quasar activity played in early galaxy evolution?

Understanding quasars
requires observations from X-ray to radio wavelengths, each spectral region
revealing different aspects of quasar emission and probing different 
regions of the active nuclei. One of the main results from the studies of 
$z\sim6$ quasars is the apparent lack of strong evolution in their 
rest-frame UV/optical and X-ray properties. Their emission-line strengths 
and UV continuum shapes are very similar to those of low-redshift quasars
\citep[e.g.][]{bar03,pen03,fan04}, the emission line ratios indicate solar or 
supersolar metallicity in emission-line regions as found in low-redshift
quasars \citep[e.g.][]{ham99,die03,fre03,mai03}, 
and the optical-to-X-ray flux ratios and X-ray continuum shapes show little 
evolution with redshift \citep[e.g.][]{str05,vig05,ste06,she05,she06}.
These measurements show that quasar accretion disks and emission-line 
regions are formed on very short time scales and their properties are not 
sensitive to the cosmic age.

However, it is not known whether this lack of evolution in high-energy 
spectral energy distributions (SEDs) at $z\sim6$ extends to the rest-frame IR
wavelength range. According to classical AGN unification 
models \citep{ant93}, the accretion disk is surrounded by a dusty torus.
Much of the emission from quasars/AGNs is re-processed by the dust 
and is re-emitted at IR wavelengths, where the quasar/AGN SEDs peak.
The hottest dust lies within a few pc and produces near-IR (NIR)
radiation, while warm and cool dust can extend to a few kpc and 
dominates the mid-IR (MIR) and far-IR (FIR) emission 
\citep[e.g.][]{pol00,nen02,sie05}. The individual contributions from AGN 
activity and star formation to the heating of warm/cool dust are poorly known
\citep[e.g.][]{wil01}. On one hand, the shapes of the MIR-to-FIR 
SEDs indicate that AGN activity may dominate the heating of the warm/cool 
dust \citep[e.g.][]{pol00,haa03}. On the other hand, the radio-to-FIR
correlation for star-forming galaxies \citep{con92} still holds for most
IR-luminous quasars out to the highest redshifts 
\citep[e.g.][]{car01,car04}, which suggests that the dust 
could be heated by starbursts in quasar host galaxies rather than quasar
central engines. However, it is generally believed that emission from hot 
dust with temperature of $\sim$1000K is directly powered by central active 
nuclei \citep[e.g.][]{rie81,pol00,haa03}, and
thus is closely related to quasar activity.

The $Spitzer\ Space\ Telescope$ \citep[$Spitzer$;][]{wer04} allows us, for the
first time, to explore the rest-frame NIR range for high-redshift
quasars and to constrain the evolution of hot dust in quasar environments.
\citet{hin06} have observed thirteen $z>4.5$ quasars using the Infrared
Array Camera \citep[IRAC;][]{faz04} and Multiband Imaging Photometer for
$Spitzer$ \citep[MIPS;][]{rie04}. They find that the SEDs of these 
high-redshift quasars at rest wavelength 0.6--4.3 $\mu$m do not significantly
differ from those of quasars with similar luminosity at low redshifts. In this 
paper, we report on $Spitzer$ observations of thirteen $z\sim6$ quasars
discovered by the SDSS. All the quasars were observed with IRAC. Ten of them 
were also observed in the MIPS 24$\mu$m band, while the 24$\mu$m fluxes of
the other three were taken from \citet{hin06}.

In $\S$2 of this paper, we present our high-redshift quasar sample and
the $Spitzer$ photometry of the thirteen quasars. In $\S$3 we show their 
rest-frame NIR SEDs and study hot dust at $z\sim6$. We calculate bolometric
luminosities and accretion rates for these quasars in $\S$4 and give the
summary and discussion in $\S$5. Throughout the paper we use $\lambda_0$ 
($\nu_0$) to denote rest-frame wavelength (frequency), and use a 
$\Lambda$-dominated flat cosmology 
with H$_0$ = 70 km s$^{-1}$ Mpc$^{-1}$, $\Omega_{m}$ = 0.3 and
$\Omega_{\Lambda}$ = 0.7 \citep[e.g.][]{spe03,spe06}.

\section{OBSERVATIONS}

\subsection{A fundamental sample of luminous $z\sim6$ quasars from the 
Sloan Digital Sky Survey}

The SDSS is the main source for high-redshift quasar discovery to date,
and has discovered more than twenty luminous 
quasars at $z>5.7$ from $\sim8000$ deg$^2$. Thirteen of them were observed
using $Spitzer$ and are included in this paper. Table 1 presents optical and 
NIR properties of the thirteen quasars. Redshifts, $M_{1450}$, $m_{1450}$, and
the photometry in $i$, $z$, and $J$ bands are mostly from the quasar 
discovery papers \citep{fan00,fan01,fan03,fan04,fan06}; $K$-band photometry of
SDSS J1044$-$0125\footnote{The naming convention for SDSS sources is SDSS
JHHMMSS.SS$\pm$DDMMSS.S, and the positions are expressed in J2000.0 
coordinates. We use SDSS JHHMM$\pm$DDMM for brevity.} is from \citet{fan00}; 
$H$- and $K'$-band photometry of SDSS J1030+0524, J1048+4637, J1148+5251 and 
J1630+4012 are from \citet{iwa04}; $J$-band photometry of SDSS J1044$-$0125 
and $H$-band photometry of SDSS J0005$-$0006, 
J0836+0054, J0840+5624 and J1044$-$0125 were carried out in November 2005
using the 6.5m MMT with SWIRC, which is a $J$- and $H$-band camera 
operating at the f/5 cassegrain focus of the MMT.
Note that the SDSS $ugriz$ photometric system \citep{fuk96} is based on the 
AB magnitude scale of \citet{oke83}, and that the photometry is reported on 
the asinh scale described in \citet{lup99}. The $JHK$ measurements are 
Vega-based magnitudes.

These $z\sim6$ quasars also have multiwavelength observations from 
radio \citep{pet03,car04,fre05,car06}, 
mm/submm \citep{pet03,pri03,ber03a,rob04}, 
to X-ray \citep{bra01,bra02,bec03,far04,sch04,she06}. 
The $Spitzer$ observations, combined with X-ray, optical, NIR, mm/submm and
radio observations, provide a fundamental sample of quasar SEDs at $z\sim6$.

\subsection{$Spitzer$ observations of thirteen $z\sim6$ quasars}

IRAC and MIPS 24$\mu$m photometry for the $z\sim6$ quasars was obtained
by our $Spitzer$ GO-1 program (3198). IRAC observations were carried out
in channels 1, 2, 3, and 4 (3.6, 4.5, 5.8 and 8.0 $\mu$m) with an exposure
time of 1000 s in each channel. Images were processed by the IRAC Basic
Calibrated Data (BCD) pipeline, and aperture photometry was performed using
customized IDP3 \citep{sch02} IDL software. We used a 6 pixel (7$\farcs3$)
target aperture radius and measured the background in an annulus from 8$-$13 
pixels. The contaminant sources within the background area were masked by 
hand. Finally aperture corrections were derived from the $Spitzer$ Tiny Tim 
simulations \citep{kri02}. The pixel-to-pixel fluctuations
in the background annuli were used to estimate the measurement uncertainties.
In addition, there is an uncertainty of 3$-$5\% in absolute calibration
\citep{rea05}.

MIPS 24$\mu$m photometry for ten quasars was obtained in our $Spitzer$ GO
program, and the other three were observed in a MIPS GTO program 
\citep{hin06}. The integration time was 1400 s for quasars with high 
background, and 1260 s for others. The background is estimated using the 
$Spitzer$ background estimator, and high background over the $Spitzer$ 
24$\mu$m passband is about 65.6 MJy/sr. Images were processed by the MIPS BCD 
pipeline and aperture photometry was performed in the same way as the 
procedure for IRAC. The target aperture radius was chosen to be 6 pixels
(15$\farcs$0) and the background
annulus from 8--13 pixels. The background fluctuations were also used to
estimate the measurement uncertainties. The uncertainty in absolute 
calibration is about 10\%.

Images at 70 $\mu$m were obtained for four of the objects in the sample (2 
from the GTO program and 2 from our GO program). 
These data were reduced with the MIPS 
Data Analysis Tool \citep{gor05}. Photometry was performed using a 35$\arcsec$
aperture with 39--65$\arcsec$ background annulus. None of the objects was 
detected at 70 $\mu$m. In order to increase the sensitivity of our 
measurements, we combined the observations into a median stacked image 
containing these four objects. While the measured noise decreased 
approximately by a factor of 2, 
no detection was achieved in the stacked image.

The observed fluxes and measurement uncertainties of the thirteen quasars 
are given in Table 2. The upper limits are constructed from the measured 
flux density in the target aperture plus two times the measurement 
uncertainty. All quasars except SDSS J0005$-$0006 were detected with 
high S/N in all IRAC bands and the MIPS 24$\mu$m band. SDSS J0005$-$0006 was 
marginally detected in the IRAC 8.0$\mu$m band, and not detected in the MIPS 
24$\mu$m band. Because of its faintness, we use small apertures on this source
(3 and 4 pixels respectively at 8.0 and 24 $\mu$m) to suppress noise; the 
measurements use appropriate aperture corrections.
Figure 1 shows the IRAC and MIPS 24$\mu$m images for SDSS J0005$-$0006 
(images in the second line) comparing with those of SDSS J0002+2550 (images 
in the first line), whose SED in the $Spitzer$ bands is consistent with 
low-redshift SED templates (see $\S$3 and Figure 2). 

\section{SPECTRAL ENERGY DISTRIBUTIONS AND HOT DUST AT HIGH REDSHIFT}

\subsection{Spectral Energy Distributions}

Figure 2 shows the SEDs of the thirteen quasars from the $Spitzer$ 
observations. Dotted lines are the average quasar SED from \citet{elv94} and 
dashed lines are the SED template of luminous SDSS quasars from \citet{ric06}.
The SED templates have been normalized at rest-frame 1450 \AA. All these
quasars have [3.6]$_{AB}-$[4.5]$_{AB}>-0.1$ ([3.6]$_{AB}$ and [4.5]$_{AB}$ are
AB magnitudes at 3.6 and 4.5 $\mu$m, respectively), which is a MIR selection 
criterion for AGN used by \citet{coo06}. The IRAC 4.5$\mu$m 
fluxes in some objects are significantly increased by strong H$\alpha$ 
emission lines. For most quasars, the continuum shapes at the wavelengths that 
the IRAC and MIPS 24$\mu$m bands cover (0.5 $\mu$m$<\lambda_0<$3.5 $\mu$m) are 
well predicted by the low-redshift SEDs. In a type I quasar, the radiation at 
$\lambda_0<1$ $\mu$m is from the accretion disk; at longer wavelengths of a 
few microns, emission from hot dust dominates over the disk emission.
Figure 2 shows that even at $z\sim6$, accretion disks and hot-dust
structures for most quasars may already have reached maturity.

However, two quasars in this sample, SDSS J0005$-$0006 ($z=5.85$) and SDSS 
J1411+1217 ($z=5.93$), stand out as 
having unusual SEDs in this wavelength range: their fluxes at 24 $\mu$m 
($\lambda_0\sim$3.5 $\mu$m) and/or 8 $\mu$m ($\lambda_0\sim$1 $\mu$m) 
significantly deviate from the low-redshift SEDs. For SDSS J1411+1217, the 
observed flux at 24 $\mu$m is 
a factor of $\sim 3$ lower than that of the standard low-redshift templates. 
SDSS J0005-0006 is more extreme: it is marginally detected at 8 $\mu$m
and is completely undetected at 24 $\mu$m, at least an order of magnitude
fainter than that predicted by the standard templates.

\subsection{Hot dust in $z\sim6$ quasars}

We use a simple model to fit SEDs to our broad-band data at rest-frame
0.15$-$3.5 $\mu$m, consisting of a power-law disk component and a hot 
dust blackbody \citep{gli06}.
\citet{gli06} find that a 1260K blackbody provides
a good description of hot dust from their NIR quasar composite spectrum.
Because only the 24$\mu$m fluxes are available to constrain hot dust
for these $z\sim6$ quasars, we fix the hot-dust temperature as 1260 K
in our model fitting. Figure 3 shows the results of this fit. The dotted 
lines show the two components, and the dashed line is the sum of the two. 
The power-law slope $\alpha$ ($f_{\nu}\sim \nu^{\alpha}$) is also given in 
the figure. Most quasars have prominent hot-dust components seen as excess
emission in the 24$\mu$m band, above the power-law disk components, while 
SDSS J0005$-$0006 and SDSS J1411+1217 do not show any hot-dust emission. The 
SED of SDSS J1411+1217 is consistent with a pure power-law over the full 
spectral range, and the fluxes of SDSS J0005$-$0006 at $\lambda_0\sim$1 
and 3.5 $\mu$m lie significantly below the power-law.
At $z\sim6$, quasar host galaxies are very young. It is possible that
the properties of dust, including its temperature, composition and 
geometry, are different in such young objects from those at lower redshift
(see $\S5$ for a detailed discussion).

Radio and mm/submm \citep[e.g.][]{ber03a,pri03,rob04,car04} observations 
have revealed that these luminous $z\sim6$ quasars have large amounts of
warm/cool dust with masses higher than $10^8M_{\sun}$. The warm/cool dust has 
temperatures from a few tens to a few hundred kelvins, while hot dust has
temperatures of $\sim1000$ K. In Figure 3 we fit the hot dust emission using 
a single-temperature blackbody, so we cannot calculate the hot-dust mass. 
However, we may estimate a lower limit to this mass by assuming that the hot 
dust radiates as a gray body. 
The dust mass $M_d$ is determined by the following relation \citep{hug97},
\begin{equation}
M_d=\frac{F_1 D_L^2}{k_0B(\nu_0,T_d)(1+z)},
\end{equation}
where $F_1$ is the observed flux density, $D_L$ is the luminosity distance,
$k_0$ is the rest-frame dust absorption coefficient, and $B(\nu_0,T_d)$ is
the Planck function at rest frequency $\nu_0$ and temperature $T_d$.
We calculate the mass absorption coefficient $k_0$ according to \citet{loe97} 
based on the extinction law of \citet{mat90}. The lower limits of hot-dust
masses are given in Figure 3.

In the analysis above we have neglected the contribution from the host 
galaxies of quasars. According to the correlation between central BH mass and 
host luminosity \citep{peng06}, the $R$-band luminosity is about $M_R=-24$
for a galaxy hosting a BH with a mass of a few $10^9$ M$_\sun$. Quasars
with similarly massive BHs in our sample (see $\S$4) are more luminous than
$M_i=-27$. Using the elliptical galaxy template of \citet{fio97}, we find
that the contribution from the host galaxies is expected to be less than 10\% 
at rest-frame 1.6 $\mu$m, the peak of the template.
We also use galaxy templates (Sbc, Scd, and Im) of \citet{col80}, and find
that the contribution from the host galaxies is less than 10\% at rest-frame 
5000--6000\AA, the peak of the optical band in the templates.

Figure 4 shows the correlation between rest-frame 4400\AA\ luminosity and 
3.5$\mu$m luminosity for low-$z$ type I quasars, compared with our 
measurements at high redshift (red points). We include PG quasars 
\citep[green points;][]{sch83} that were observed at 3.7$\mu$m \citep{neu87}, 
4.8$\mu$m, 6.7$\mu$m or 7.3$\mu$m \citep{haa00,haa03}. We also include SDSS 
quasars (blue points) in the $Spitzer$ Extragalactic First Look Survey and 
three $Spitzer$ Wide-Area Infrared Extragalactic Survey areas \citep{ric06}.
Cyan points present $z\sim5$ quasars from \citet{hin06}. Dashed lines show
the best linear fit and its $3\sigma$ range. The correlation given in Figure 4
suggests that the dust emission at 3.5 $\mu$m in quasars is heated directly 
by central engines \citep[e.g.][]{rie81,pol00,haa03}. The NIR-to-optical
flux ratios for most $z\sim6$ quasars follow those at low redshift.
However, SDSS J0005$-$0006 and SDSS J1411+1217 lie significantly below the 
linear fit. SDSS J1411+1217 is as IR-weak (IR-weak in this paper means small 
IR-to-optical ratios rather than weak absolute IR fluxes) as the most extreme 
examples at low redshift, while SDSS J0005$-$0006 is the most IR-weak
object at any redshift in the figure. It is worth noting the quasars in 
Figure 4 are all typical type I quasars, and we do not consider other types
of AGNs. For example, in Seyfert galaxies host galaxies may dominate the
radiation at optical and IR wavelengths; while in blazars the optical emission 
could be boosted due to beaming effects, resulting in small IR-to-optical 
ratios.

\subsection{Notes on individual objects}

{\bf SDSS J000552.34$-$000655.8 ($z=5.85$).} SDSS J0005$-$0006 is selected 
from the SDSS Southern Survey, a deep survey repeatedly imaging the Fall 
Celestial Equatorial Stripe in the Southern Galactic Cap. It is the
faintest quasar ($z_{AB}=20.54$) in our sample. It also has the narrowest
Ly$\alpha$ emission line of $z\sim6$ quasars \citep{fan04}. This object 
is marginally detected in the IRAC 8.0$\mu$m band and is not detected in the 
MIPS 24$\mu$m band. Due to large measurement uncertainties, its SED is poorly 
constrained at both wavelengths. Further deep $Spitzer$ observations,
such as the IRAC 8.0$\mu$m, the Infrared Spectrograph (IRS)
Peak-Up Imaging 16$\mu$m and the MIPS 24$\mu$m photometry, are needed to 
place a strong constraint on its IR SED.

{\bf SDSS J083643.85+005453.3 ($z=5.82$).} SDSS J0836+0054 is the only known
quasar at $z>5.7$ detected by the Faint Images of the Radio Sky at Twenty-cm
\citep[FIRST;][]{bec95}. Its radio flux is variable. The flux at 1.4 GHz 
changed from 1.11$\pm$0.15 mJy as measured by FIRST to 1.75$\pm$0.04 mJy of 
\citet{pet03}, a change of a factor of $\sim60$\%, and the flux at 5 GHz 
changed 
from 0.58$\pm$0.06 mJy of \citet{pet03} to 0.34$\pm$0.06 mJy of \citet{fre05}
by a factor of $\sim70$\%. SDSS J0836+0054 has a strong and broad Ly$\alpha$
emission line \citep{fan01} and a relatively weak MIPS 24$\mu$m flux. 

{\bf SDSS J104433.04$-$012502.2 ($z=5.74$).} 
SDSS J1044$-$0125 is the first quasar discovered at $z>5.7$ \citep{fan00}. 
It has weak X-ray emission \citep{bra01} and was
confirmed to be a broad absorption line (BAL) quasar \citep{goo01,djo01}. Submm
observations at 850 $\mu$m reveal the existence of $\sim4\times$10$^8$ 
M$_{\sun}$ of cool dust in this object \citep{pri03}. It is also bright at 
24$\mu$m, and thus has a large amount of hot dust.

{\bf SDSS J104845.05+463718.3 ($z=6.20$).} SDSS J1048+4637 is the most distant
known BAL quasar \citep{maio04}. It has been marginally detected
at 1.4 GHz \citep{car04}, and not detected at 450 and 850 $\mu$m 
\citep{rob04}. \citet{ber03a} detected it at 250 GHz and estimated a cool dust 
mass of $\sim4\times$10$^8$ M$_{\sun}$ in this object. It was observed with
IRS Peak-Up Imaging at 16$\mu$m \citep{cha04}, and the updated 16$\mu$m flux
is 0.49 mJy.

{\bf SDSS J114816.64+525150.2 ($z=6.42$).} SDSS J1148+5251 is the most distant 
quasar known. It has strong MIPS 24$\mu$m output, indicating the existence of 
prominent hot dust. It was also
detected at 450 and 850 $\mu$m \citep{rob04}, 250 GHz \citep{ber03a} and 1.4 
GHz \citep{car04}. These observations show that it has copious cool
dust with a mass of $5-7\times$10$^8$ M$_{\sun}$. CO observations 
\citep{ber03b,wal03,wal04} reveal the presence of $\sim2\times$10$^{10}$ 
M$_{\sun}$ of molecular gas in this object. It was also observed with
IRS Peak-Up Imaging at 16$\mu$m \citep{cha04}, and its updated 16$\mu$m flux
is 0.84 mJy. \citet{mah05} discovered a very faint quasar
RD J114816.2+525339 with $z_{AB}=23.0$ at $z=5.7$ near SDSS J1148+5251. We 
detected this faint source with IRAC at 3.6 and 4.5 $\mu$m with flux densities
of 0.013$\pm$0.002 and 0.013$\pm$0.003 mJy, respectively, which are consistent
with standard low-$z$ SED templates.

{\bf SDSS J141111.29+121737.4 ($z=5.93$).} SDSS J1411+1217 has a very weak 
MIPS 24$\mu$m flux, and its SED at 1 $\mu$m$<\lambda_0<3.5$ $\mu$m can be 
fitted with a pure power-law. It also has a narrow Ly$\alpha$ emission line
\citep{fan04}.

\section{BOLOMETRIC LUMINOSITIES AND ACCRETION RATES}

Figure 5 presents the SEDs (filled circles) of the thirteen $z\sim6$ quasars 
from X-ray to radio. The data other than our $Spitzer$ observations are taken 
from the literature mentioned in $\S2.1$. Filled circles with downward arrows 
are 2$\sigma$ upper limits. Dotted lines are the average quasar SED from 
\citet{elv94} and dashed lines are the SED template of luminous SDSS quasars 
from \citet{ric06}. The templates have been normalized at IRAC 3.6 $\mu$m. 
Due to the anticorrelation between X-ray emission and optical luminosity in 
AGNs \citep[e.g.][]{str05,ste06}, the SED template of \citet{ric06} at X-ray
band is well below that of \citet{elv94} and the X-ray fluxes of the most
luminous $z\sim6$ quasars are often smaller than the \citet{ric06} template.
After correcting for this effect, the X-ray emission
from these $z\sim6$ quasars is consistent with the low-$z$ SED templates
\citep{she06}. To calculate bolometric luminosities, we first determine 
the full SED for each quasar. The SED between any two adjacent data points is
interpolated using the \citet{ric06} mean SED, and the SED beyond the leftmost
or rightmost points is directly scaled to this point using the mean SED.
Table 3 gives the bolometric luminosities as well as optical and IR 
luminosities. Column 5 is the bolometric correction from rest-frame B band for 
all the quasars. The mean correction and standard deviation are $9.1\pm2.2$, 
consistent with $11.8\pm4.3$ from \citet{elv94} and $10.4\pm2.5$ from
\citet{ric06}.

We estimate central BH masses for the quasars that have been spectroscopically 
observed at rest-frame UV/optical wavelength \citep[e.g.][]{goo01,bar03} using 
the BH mass scaling relations \citep{mcl04,ves06}. The BH mass of SDSS 
J0836+0054 is taken from Pentericci et al. (2006, in prep.) based on the 
relation of \citet{mcl04}. \citet{iwa04} measured 
the MgII emission lines for four of the quasars. Due to low S/N, their 
measured MgII width for SDSS J1148+5251 is lower than that determined by 
\citet{bar03} by a factor of $\sim2$, which results in a factor of 4 lower 
estimated BH 
mass. We thus did not use the measurements of \citet{iwa04}. The derived BH 
masses and Eddington luminosity ratios are given in Table 3. These luminous 
$z\sim6$ quasars have supermassive BHs with masses of a few $10^9$ M$_{\sun}$ 
and Eddington ratios of order unity, comparable to quasars with similar
luminosities at lower redshift \citep[e.g.][]{mcl04,ves04,kol06}.

\section{SUMMARY AND DISCUSSION}

We have carried out IRAC and MIPS 24$\mu$m photometry for thirteen $z\sim6$ 
quasars. All the quasars except SDSS J0005$-$0006
were detected with high S/N in the IRAC and MIPS 24$\mu$m bands, while SDSS 
J0005$-$0006 was marginally detected in the IRAC 8.0$\mu$m band, and not 
detected in the MIPS 24$\mu$m band. The sample of the quasars is used
to study the properties of IR SEDs and hot dust in high-redshift quasars. We 
find that the SEDs of most quasars follow low-redshift SEDs at the probed
wavelengths. However, two quasars, SDSS J0005$-$0006 and SDSS J1411+1217, have 
unusual SEDs that lie significantly below the prediction of low-redshift SED
templates at 24 $\mu$m and/or 8 $\mu$m, showing a strong IR deficit.
A simple model shows that most of the quasars have substantial hot dust, 
while the two IR-weak quasars do not show any hot-dust emission. We combine 
the $Spitzer$
observations with X-ray, UV/optical, mm/submm and radio observations to
determine bolometric luminosities for the high-redshift quasars. We find that
these quasars have Eddington ratios of order unity.

It has been 
revealed that $z\sim6$ quasars show a lack of evolution in their
rest-frame UV/optical and X-ray SEDs. Our $Spitzer$ observations show that,
for most $z\sim6$ quasars, NIR-to-MIR SEDs also do not differ significantly
from those at low redshift. This suggests that accretion disks, emission-line 
regions and dust structures in most high-redshift quasars have reached 
maturity very early on. However, we found two quasars with a strong 
IR deficit. As shown in Figure 4, NIR-weak Type I quasars like SDSS 
J1411+1217 are very rare; and Type I quasars with extremely weak NIR fluxes
like SDSS J0005$-$0006 are not found in a large sample at low redshift, but 
exist in a small sample of thirteen quasars at $z\sim6$, suggesting that
some quasars at high redshift may have different dust properties.

In the local universe, most dust in the ISM is produced by low and
intermediate-mass AGB stars, which develop 0.5$-$1 Gyr after the initial 
starburst. At $z\sim 6$, the age of the universe is less than 1 Gyr, so 
quasar host galaxies are very young, with their first star formation likely 
to have occurred less than half a Gyr earlier. The two NIR-weak quasars SDSS 
J0005$-$0006 and SDSS 1411+1217 could be too young to have formed dust tori 
around them, or perhaps, dust properties in such young systems
differ from those of lower-redshift quasars. Thus the unusual IR SEDs of
the two quasars may be a reflection of different dust properties in very young 
host galaxies at high redshift. \citet{mai04} found that the dust extinction
curve in quasar SDSS J1048+4836 ($z=6.20$) is different from that observed at
$z<4$ (which is SMC-like, \citet{hop04}), implying a grain size distribution 
more similar to that expected from dust produced by supernovae.
While other dust production mechanisms \citep[e.g.][]{elv02} are also 
possible, it is currently unknown what their emission spectra are. 
SDSS J0005$-$0006 is especially interesting. If it has 
no hot dust, the observed SED at 1 $\mu$m$<\lambda_0<3.5$ $\mu$m comes
from the disk only. This provides a strict constraint on disk models.

Hot dust in SDSS J0005$-$0006 and SDSS J1411+1217 could also be hidden by 
dust tori. According to AGN unification models, the accretion disk is 
surrounded by a dust torus. The hottest dust, with temperature more than 
1000K, lies within a few pc, while cool dust with temperature of a few tens 
of kelvins can extend to a few kpc. As the angle between the torus axis and 
the line of sight increases, more hot dust could be hidden by the cool dust 
torus. Thus the hot-dust radiation in the NIR would be suppressed, and the
$\lambda_0\sim1\mu$m dip could be shifted toward longer wavelengths
\citep{haa03}. Intriguingly, the emission line widths in the two NIR-weak 
quasars are the narrowest in our sample of $z\sim 6$ quasars. This suggests 
that the weak NIR SEDs in two sources could be caused by an obscuration 
with hot dust hidden by cooler dust tori. On the other hand, both objects 
show normal UV continuum and UV-to-X-ray flux ratio, and are thus 
not type II obscured objects in the normal sense, but could be intermediate 
objects in which the hot dust contribution is somewhat reduced.
Further deep $Spitzer$ observations are needed to constrain the SED shapes at 
$1\mu m<\lambda_0<4\mu m$, and NIR spectra are also needed
to detect possible narrow line components.

\acknowledgments

We acknowledge support from NSF grant AST-0307384, a Sloan Research Fellowship
and a Packard Fellowship for Science and Engineering (L.J., X.F., M.V.). 
We acknowledge support from NASA LTSA grant NAG5-13035 (WNB and DPS). We 
thank the SWIRC instrument team and MMT staff for their expert help in 
preparing and carrying out the SWIRC observing run.

This work is based on observations made with the Spitzer Space Telescope, 
which is operated by the Jet Propulsion Laboratory, California Institute of 
Technology under a contract (3198) with NASA. Support for this work was 
provided by NASA through an award issued by JPL/Caltech.

Funding for the SDSS and SDSS-II has been provided by the Alfred P. Sloan
Foundation, the Participating Institutions, the National Science Foundation,
the U.S. Department of Energy, the National Aeronautics and Space
Administration, the Japanese Monbukagakusho, the Max Planck Society, and the
Higher Education Funding Council for England.
The SDSS Web Site is http://www.sdss.org/.
The SDSS is managed by the Astrophysical Research Consortium for the
Participating Institutions. The Participating Institutions are the American
Museum of Natural History, Astrophysical Institute Potsdam, University of
Basel, Cambridge University, Case Western Reserve University, University of
Chicago, Drexel University, Fermilab, the Institute for Advanced Study, the
Japan Participation Group, Johns Hopkins University, the Joint Institute for
Nuclear Astrophysics, the Kavli Institute for Particle Astrophysics and
Cosmology, the Korean Scientist Group, the Chinese Academy of Sciences
(LAMOST), Los Alamos National Laboratory, the Max-Planck-Institute for
Astronomy (MPIA), the Max-Planck-Institute for Astrophysics (MPA), New Mexico
State University, Ohio State University, University of Pittsburgh, University
of Portsmouth, Princeton University, the United States Naval Observatory, and
the University of Washington.

\clearpage
\begin{figure}
\plotone{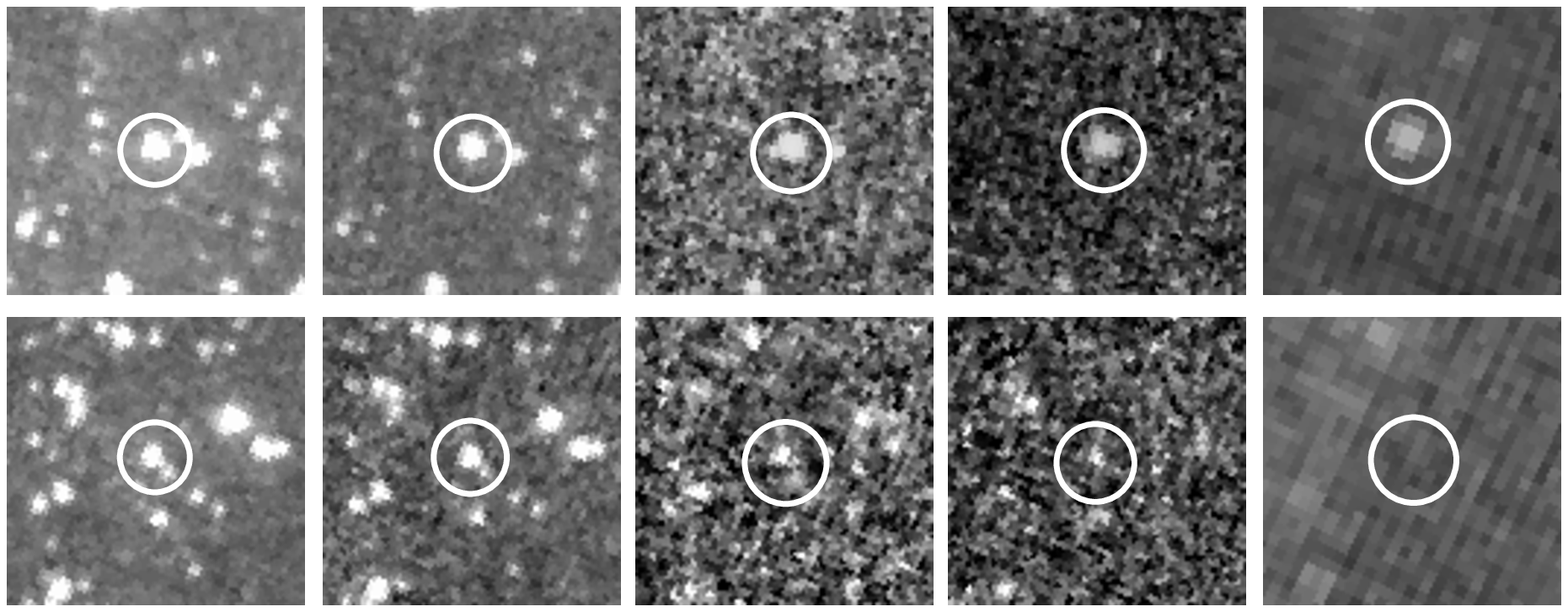}
\caption{IRAC and MIPS 24$\mu$m images of SDSS J0005$-$0006 (images in the 
second line) compared with those of SDSS J0002+2550 (images in the first 
line), whose SED in the $Spitzer$ bands is consistent with low-redshift 
SED templates (see $\S$3 and Figure 2). The images from left to right 
correspond to IRAC channels 1, 2, 3 and 4, and the MIPS 24$\mu$m band. The 
size of the images is $1'\times 1'$. SDSS J0005$-$0006 was marginally detected 
in the IRAC 8.0$\mu$m band, and is not detected in the MIPS 24$\mu$m band.}
\end{figure}

\clearpage
\begin{figure}
\epsscale{0.65}
\plotone{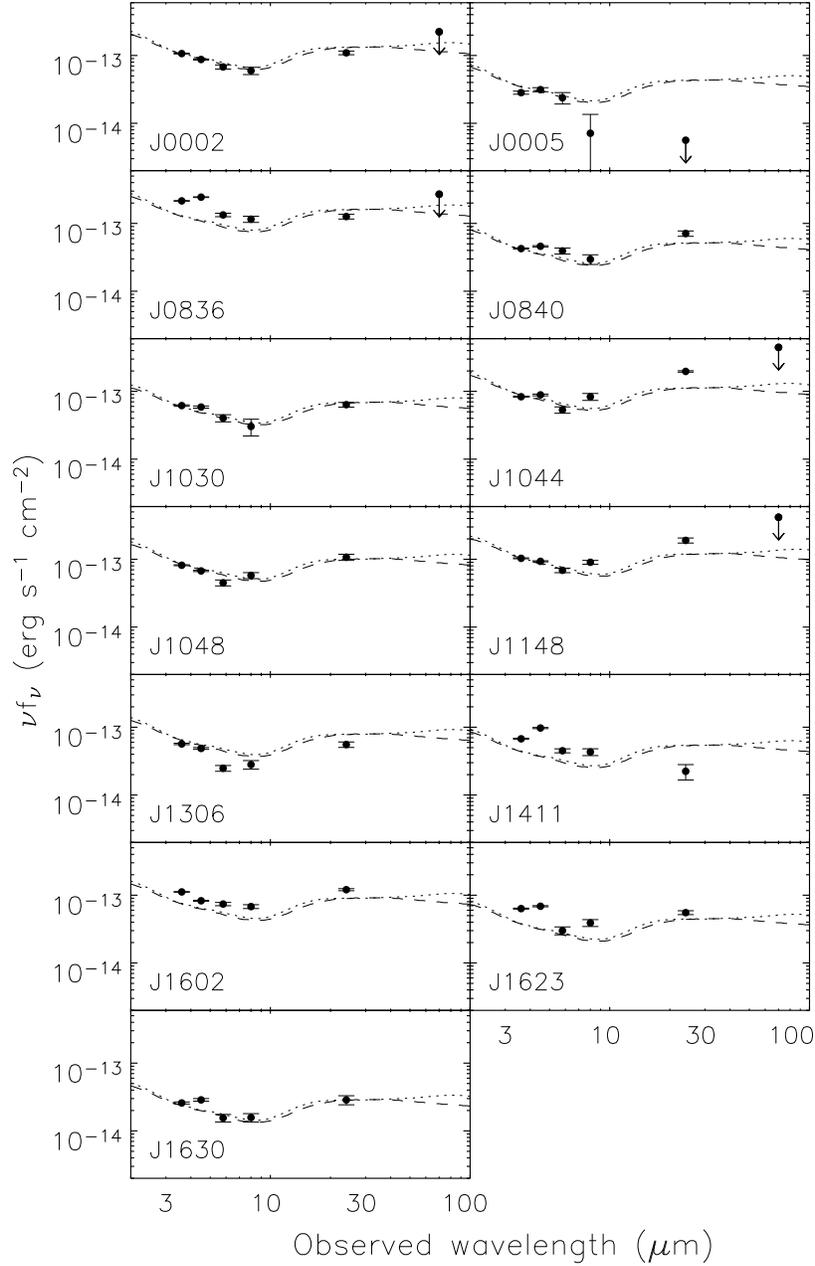}
\caption{SEDs of the thirteen quasars from the $Spitzer$ observations. The
dotted and dashed lines are the average SEDs of type I low-redshift quasars 
from \citet{elv94} and \citet{ric06}, respectively, and have been normalized
at rest-frame 1450 \AA. Filled circles with downward arrows are 2$\sigma$ 
upper limits. The measurement uncertainties are also given in the figure.}
\end{figure}

\clearpage
\begin{figure}
\plotone{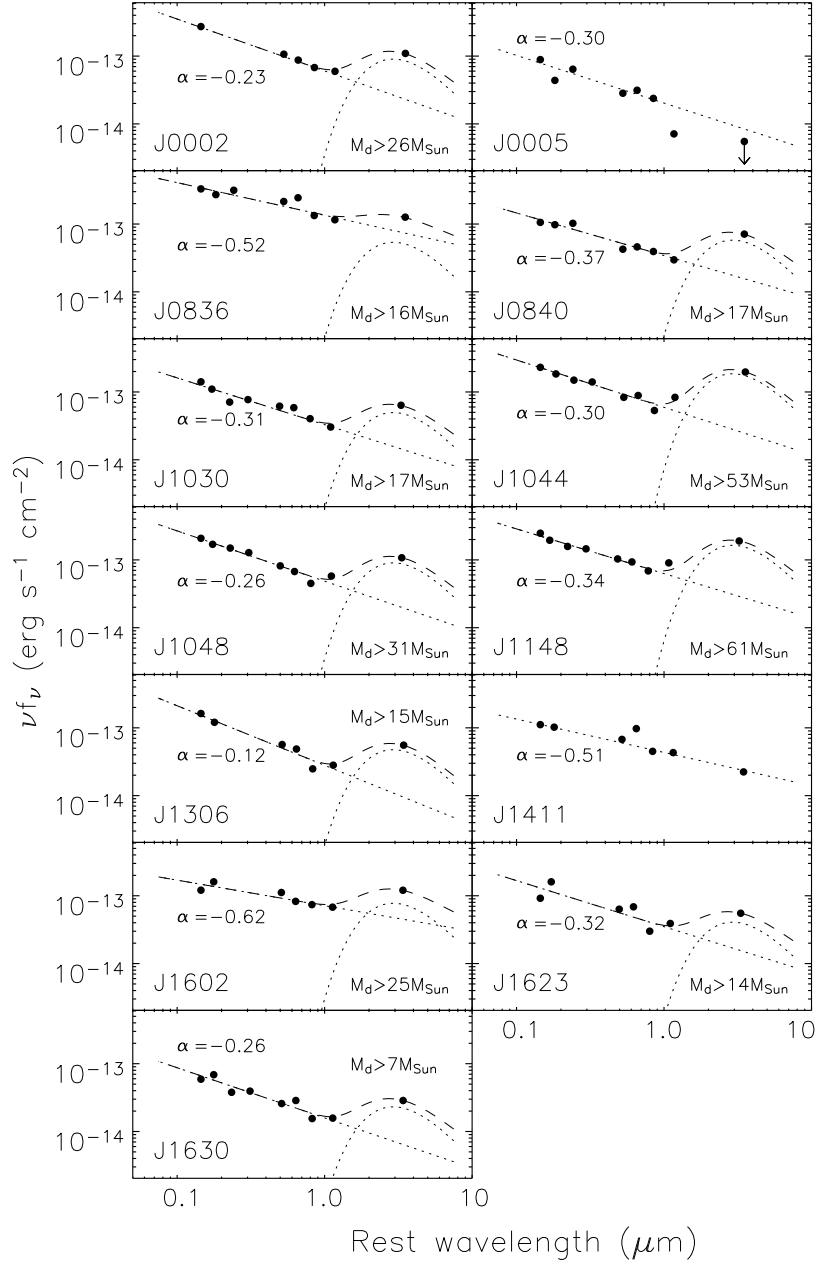}
\caption{A simple model fitting to the high-redshift quasar SEDs at rest-frame
$0.15-3.5$ $\mu$m. The dotted lines in each panel show a power-law disk 
component and a blackbody component of hot dust, and the dashed line is the 
sum of the two. The power-law slope $\alpha$ and the lower limit of hot-dust
mass $M_d$ for each object are also given in the figure.}
\end{figure}

\clearpage
\begin{figure}
\epsscale{1.0}
\plotone{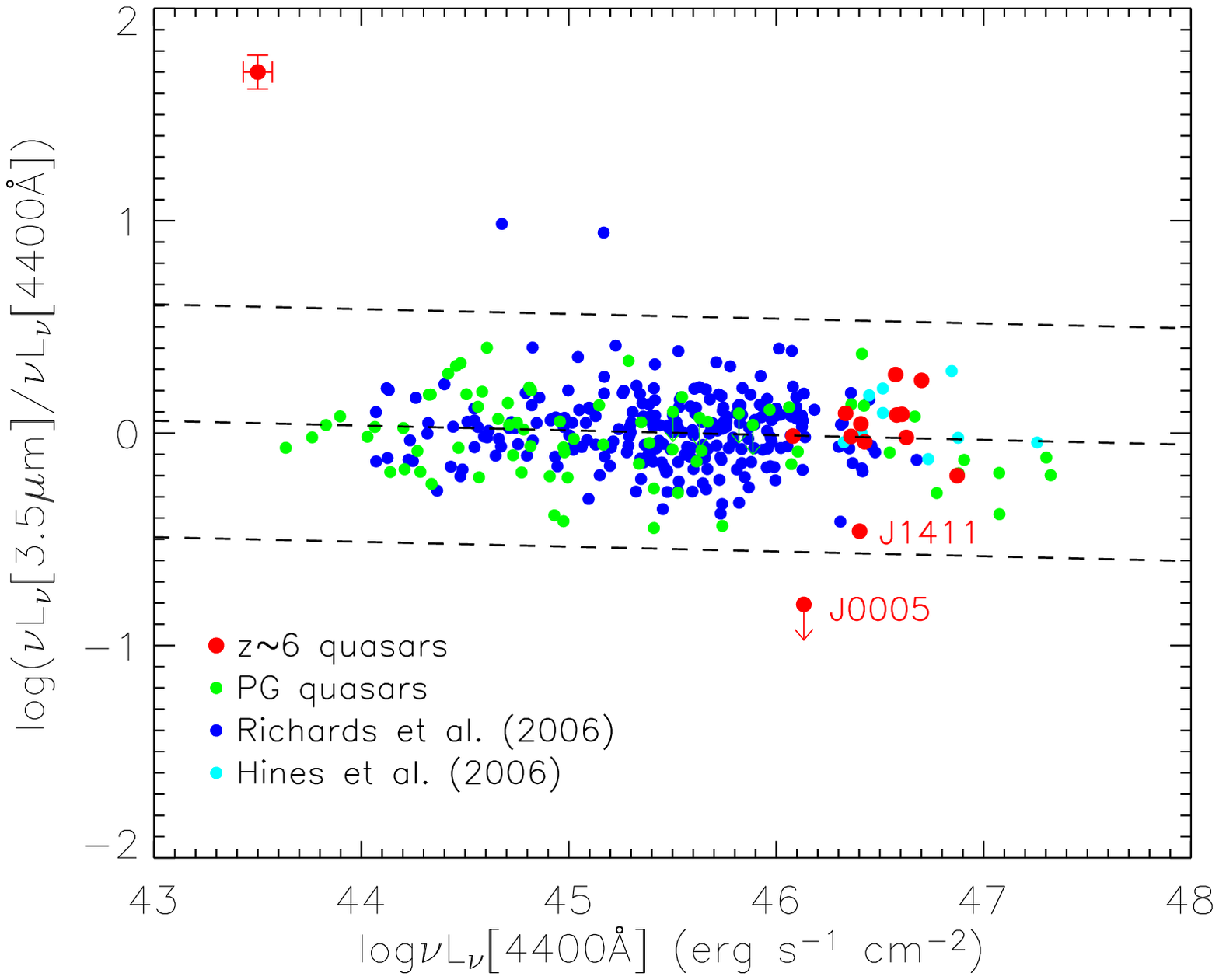}
\caption{Correlation between rest-frame 4400\AA\ luminosity and 3.5$\mu$m 
luminosity for type I quasars. Red points are our $z\sim6$ quasars.
Green points are PG quasars \citep{sch83} that were observed at 3.7$\mu$m
\citep{neu87}, 4.8$\mu$m, 6.7$\mu$m or 7.3$\mu$m \citep{haa00,haa03}.
Cyan points are $z\sim5$ quasars from \citet{hin06}. Filled circles with 
downward arrows are 2$\sigma$ upper limits. Dashed lines show
the best linear fit and its $3\sigma$ range. Typical errors for $z\sim6$
quasars are given in the upper left corner. SDSS J1411+1217 is as IR-weak
as the most extreme examples at low redshift, while SDSS J0005$-$0006 is the 
most IR-weak quasar at any redshift in the figure.}
\end{figure}

\clearpage
\begin{figure}
\epsscale{0.65}
\plotone{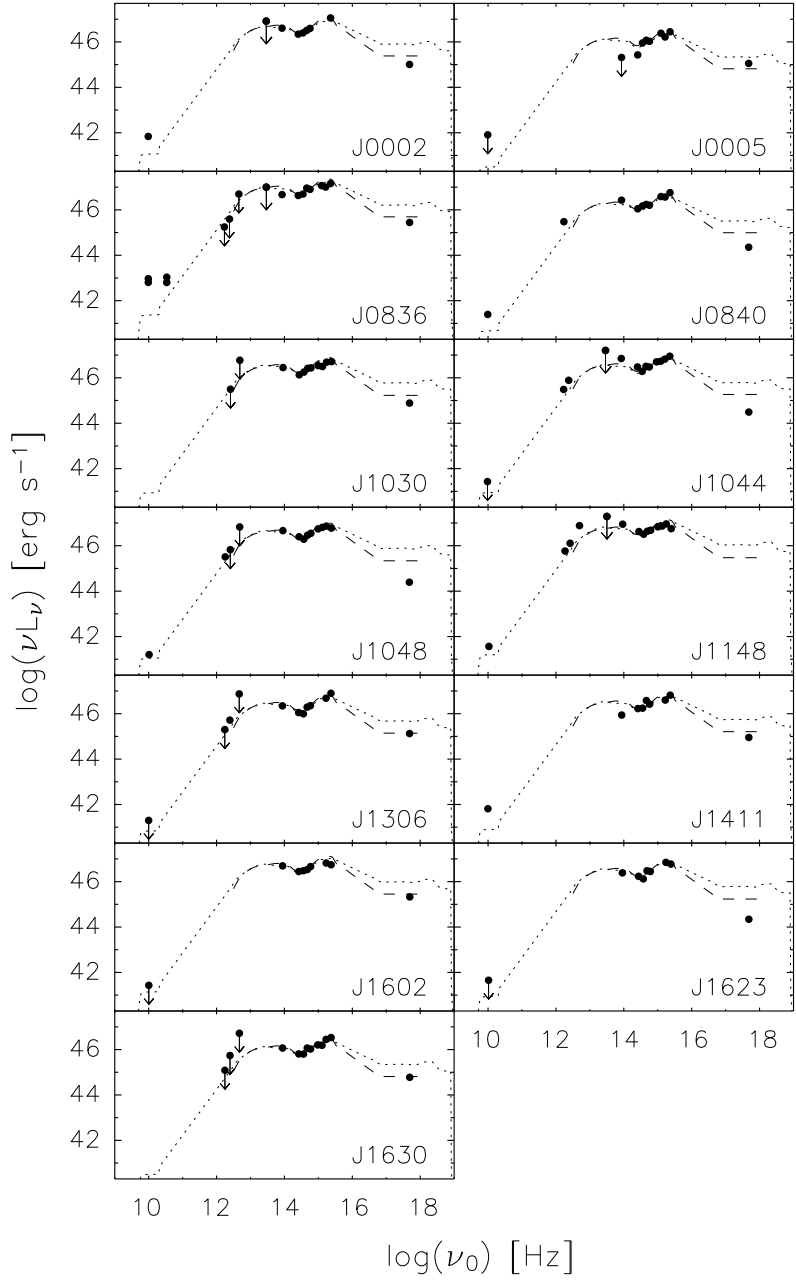}
\caption{SEDs of the thirteen quasars from X-ray to radio. The data other than
our $Spitzer$ observations are taken from the literature mentioned in $\S2.1$.
Filled circles with downward arrows are 2$\sigma$ upper limits.
The dotted and dashed lines are the average SEDs of low-redshift quasars
from \citet{elv94} and \citet{ric06}, respectively, and have been normalized
at IRAC 3.6 $\mu$m ($\lambda_0\sim5000$\AA).}
\end{figure}

\clearpage
\begin{deluxetable}{cccccccccc}
\rotate
\tablecaption{Optical and NIR properties of the thirteen quasars}
\tablewidth{0pt}
\tablehead{
\colhead{Quasar (SDSS)} & \colhead{redshift} & \colhead{$M_{1450}$} & 
\colhead{$m_{1450}$}    & \colhead{$i$}      & \colhead{$z$}        & 
\colhead{$J$}           & \colhead{$H$}      & \colhead{$K'$ (or $K$)} & 
\colhead{Ref.}}
\startdata
J000239.39+255034.8   & 5.80 & $-27.7$ & 19.02 & 21.47 & 18.99 & $\ldots$ & $\ldots$ & $\ldots$ &  4 \\
J000552.34$-$000655.8 & 5.85 & $-26.2$ & 20.23 & 23.40 & 20.54 & 19.87    & 18.68    & $\ldots$ &  4 \\
J083643.85+005453.3   & 5.82 & $-27.9$ & 18.81 & 21.04 & 18.74 & 17.89    & 16.95    & $\ldots$ &  2 \\
J084035.09+562419.9   & 5.85 & $-26.9$ & 20.04 & 22.43 & 19.76 & 19.00    & 18.17    & $\ldots$ &  5 \\
J103027.10+052455.0   & 6.28 & $-27.2$ & 19.66 & 23.23 & 20.05 & 18.87    & 18.57    & 17.67    &  2,6 \\
J104433.04$-$012502.2 & 5.74 & $-27.5$ & 19.21 & 21.81 & 19.23 & 18.31    & 17.76    & 17.02($K$)& 1 \\
J104845.05+463718.3   & 6.20 & $-27.6$ & 19.25 & 22.38 & 19.86 & 18.40    & 17.83    & 17.12    &  3,6 \\
J114816.64+525150.2   & 6.42 & $-27.8$ & 19.03 & 23.86 & 20.12 & 18.25    & 17.62    & 16.98    &  3,6 \\
J130608.26+035626.3   & 5.99 & $-26.9$ & 19.55 & 22.58 & 19.47 & 18.77    & $\ldots$ & $\ldots$ &  2 \\
J141111.29+121737.4   & 5.93 & $-26.8$ & 19.97 & 22.85 & 19.65 & 18.95    & $\ldots$ & $\ldots$ &  4 \\
J160254.00+422825.0   & 6.07 & $-26.8$ & 19.86 & 22.78 & 19.89 & 18.46    & $\ldots$ & $\ldots$ &  4 \\
J162331.81+311200.5   & 6.22 & $-26.7$ & 20.13 & 24.52 & 20.09 & 19.15    & $\ldots$ & $\ldots$ &  4 \\
J163033.90+401209.6   & 6.05 & $-26.2$ & 20.64 & 23.38 & 20.42 & 19.38    & 19.18    & 18.40    &  3,6 \\
\enddata
\tablecomments{References (1) \citet{fan00}; (2) \citet{fan01}; (3) 
\citet{fan03}; (4) \citet{fan04}; (5) \citet{fan06}; (6) \citet{iwa04}.
Redshifts, $M_{1450}$, $m_{1450}$, and
the photometry in $i$, $z$, and $J$ bands are mostly from the quasar
discovery papers \citep{fan00,fan01,fan03,fan04,fan06}; $K$ photometry of
SDSS J1044$-$0125 is from \citet{fan00}; $H$ and $K'$ photometry of SDSS
J1030+0524, J1048+4637, J1148+5251 and J1630+4012 are from \citet{iwa04};
$J$ photometry of SDSS J1044$-$0125 and $H$ photometry of SDSS J0005$-$0006,
J0836+0054, J0840+5624 and J1044$-$0125 were carried out in November 2005
using the 6.5m MMT with SWIRC. The SDSS photometry of $i$ and $z$ is on
 the AB system; $J$, $H$, $K$ and $K'$ are on a Vega-based system.}
\end{deluxetable}

\clearpage
\begin{deluxetable}{ccccccc}
\rotate
\tablecaption{$Spitzer$ photometry of the thirteen quasars}
\tablewidth{0pt}
\tablehead{
\colhead{Quasar (SDSS)} & \colhead{3.6$\mu$m(mJy)} & \colhead{4.5$\mu$m(mJy)} &
\colhead{5.8$\mu$m(mJy)} & \colhead{8.0$\mu$m(mJy)} & \colhead{24$\mu$m(mJy)} &
\colhead{70$\mu$m(mJy)\tablenotemark{a}}}
\startdata
J0002+2550   & 0.128$\pm$0.002 & 0.131$\pm$0.002 & 0.131$\pm$0.009 & 0.159$\pm$0.019 & 0.876$\pm$0.055 & 5.21 \\
J0005$-$0006 & 0.034$\pm$0.002 & 0.047$\pm$0.003 & 0.046$\pm$0.009 & 0.019$\pm$0.017\tablenotemark{b} & 0.004$\pm$0.022\tablenotemark{b} & $\ldots$ \\
J0836+0054   & 0.258$\pm$0.003 & 0.366$\pm$0.004 & 0.258$\pm$0.015 & 0.308$\pm$0.032 & 1.010$\pm$0.080\tablenotemark{c} & 6.29 \\
J0840+5624   & 0.051$\pm$0.001 & 0.069$\pm$0.002 & 0.076$\pm$0.008 & 0.079$\pm$0.013 & 0.568$\pm$0.051 & $\ldots$ \\
J1030+0524   & 0.074$\pm$0.002 & 0.088$\pm$0.003 & 0.078$\pm$0.009 & 0.081$\pm$0.023 & 0.509$\pm$0.041 & $\ldots$ \\
J1044$-$0125 & 0.100$\pm$0.002 & 0.133$\pm$0.004 & 0.103$\pm$0.011 & 0.222$\pm$0.026 & 1.575$\pm$0.038 & 10.4 \\
J1048+4637   & 0.098$\pm$0.002 & 0.101$\pm$0.002 & 0.087$\pm$0.009 & 0.154$\pm$0.016 & 0.860$\pm$0.090\tablenotemark{c} & $\ldots$ \\
J1148+5251   & 0.124$\pm$0.002 & 0.140$\pm$0.003 & 0.133$\pm$0.010 & 0.241$\pm$0.016 & 1.520$\pm$0.130\tablenotemark{c} & 9.73 \\
J1306+0356   & 0.068$\pm$0.002 & 0.073$\pm$0.002 & 0.048$\pm$0.005 & 0.075$\pm$0.011 & 0.444$\pm$0.041 & $\ldots$ \\
J1411+1217   & 0.081$\pm$0.002 & 0.146$\pm$0.003 & 0.087$\pm$0.006 & 0.115$\pm$0.013 & 0.179$\pm$0.046 & $\ldots$ \\
J1602+4228   & 0.134$\pm$0.001 & 0.124$\pm$0.002 & 0.143$\pm$0.008 & 0.181$\pm$0.012 & 0.964$\pm$0.034 & $\ldots$ \\
J1623+3112   & 0.076$\pm$0.001 & 0.103$\pm$0.002 & 0.058$\pm$0.008 & 0.104$\pm$0.012 & 0.442$\pm$0.029 & $\ldots$ \\
J1630+4012   & 0.031$\pm$0.001 & 0.043$\pm$0.002 & 0.030$\pm$0.004 & 0.042$\pm$0.006 & 0.229$\pm$0.035 & $\ldots$ \\
\enddata
\tablecomments{Errors given in this table are measurement uncertainties 
  only. The absolute calibration uncertainty for IRAC is 3--5\%, and
  for MIPS is about 10\%.}
\tablenotetext{a}{2$\sigma$ upper limits.}
\tablenotetext{b}{Using smaller apertures than others for photometry; 
  see $\S$2.2. }
\tablenotetext{c}{From \citet{hin06}.}
\end{deluxetable}

\clearpage
\begin{table}
\caption{Optical, IR and bolometric luminosities for the thirteen quasars}
\begin{tabular}{ccccccc}
\tableline\tableline
 Quasar (SDSS) & $L_{Bol}$\tablenotemark{a} & $L_{Opt}$\tablenotemark{b} & $L_{IR}$\tablenotemark{c} & $L_{Bol}/\nu L_{\nu}$(4400\AA) & $M_{BH}$($10^9 M_{\sun}$) & $L_{Bol}/L_{Edd}$ \\
\tableline
J0002+2550    & 47.57 & 47.19 & 47.18 & 8.5 & $\ldots$ & $\ldots$ \\
J0005$-$0006  & 46.94 & 46.67 & 45.96 & 6.0 & $\ldots$ & $\ldots$ \\
J0836+0054    & 47.72 & 47.36 & 47.28 & 7.0 & 6.8      & 0.54     \\
J0840+5624    & 47.34 & 46.81 & 47.11 & 9.9 & $\ldots$ & $\ldots$ \\
J1030+0524    & 47.37 & 46.91 & 47.04 & 8.7 & $\ldots$ & $\ldots$ \\
J1044$-$0125  & 47.63 & 47.06 & 47.43 & 11.3 & 6.4      & 0.53     \\
J1048+4637    & 47.55 & 47.08 & 47.29 & 9.1 & $\ldots$ & $\ldots$ \\
J1148+5251    & 47.85 & 47.19 & 47.68 & 14.3 & 5.6      & 1.01     \\
J1306+0356    & 47.40 & 46.94 & 47.04 & 10.3 & $\ldots$ & $\ldots$ \\
J1411+1217    & 47.20 & 46.91 & 46.61 & 6.3 & $\ldots$ & $\ldots$ \\
J1602+4228    & 47.59 & 47.13 & 47.27 & 9.6 & $\ldots$ & $\ldots$ \\
J1623+3112    & 47.33 & 46.98 & 46.97 & 7.6 & 3.4      & 0.50     \\
J1630+4012    & 47.06 & 46.57 & 46.75 & 9.8 & $\ldots$ & $\ldots$ \\
\tableline
\end{tabular}
\tablenotetext{a}{Bolometric luminosity (3cm to 10keV) in log(ergs/s).}
\tablenotetext{b}{Optical luminosity (0.1$\mu$m to 1.0$\mu$m) in log(ergs/s).}
\tablenotetext{c}{IR luminosity (1.0$\mu$m to 100.0$\mu$m) in log(ergs/s).}
\end{table}

\end{document}